\documentclass[12pt,preprint]{aastex}

\newcommand{\etal  }{{et al.} }
\newcommand{\msun}{\thinspace M_\odot}
\newcommand{\mj}{\thinspace M_J}
  
\newcommand{\vect}[1]{\mbox{\boldmath$#1$}}

\def\lesssim{\mathrel{\hbox{\rlap{\hbox{\lower4pt\hbox{$\sim$}}}\hbox{$<$}}}}
\def\gtrsim{\mathrel{\hbox{\rlap{\hbox{\lower4pt\hbox{$\sim$}}}\hbox{$>$}}}}
 
\newcommand{\rh  }{r_{\rm H}} 
\newcommand{\km  }{\, \rm km\,s^{-1} }

\newcommand{\dfrac}[2]{{\displaystyle \frac{#1}{#2}}  }

\shorttitle{Outflows driven by Giant Protoplanets}
\shortauthors{Machida  \etal 2006}

\begin{document}

\title{Outflows driven by Giant Protoplanets}

\author{Masahiro N. Machida\altaffilmark{1} and Shu-ichiro Inutsuka\altaffilmark{1}, and Tomoaki Matsumoto\altaffilmark{2}} 

\altaffiltext{1}{Department of Physics, Graduate School of Science, Kyoto University, Sakyo-ku, Kyoto 606-8502, Japan; machidam@scphys.kyoto-u.ac.jp, inutsuka@tap.scphys.kyoto-u.ac.jp}
\altaffiltext{2}{Faculty of Humanity and Environment, Hosei University, Fujimi, Chiyoda-ku, Tokyo 102-8160, Japan; matsu@i.hosei.ac.jp}

\begin{abstract}
 We investigate outflows driven by a giant protoplanet using three-dimensional MHD nested grid simulations.
 We consider a local region around the protoplanet in the protoplanetary disk, and calculate three models:  unmagnetized disk model,  magnetized disk model having magnetic field azimuthally parallel to the disk, and  magnetic field perpendicular to the disk.
 Outflows with velocities, at least, $\sim$10 $\km$  are driven by the protoplanets in both magnetized disk models, while outflow does not appear in unmagnetized disk model.
 Tube-like outflows along the azimuthal direction of the protoplanetary disk appear in the model with magnetic field being parallel to the disk.
 In this model, the magnetically dominated regions (i.e., density gap) are clearly contrasted from other regions and spiral waves appear near the protoplanet.
 On the other hand, in the model with magnetic field being perpendicular to the disk, outflows are driven by a protoplanet with cone-like structure just as seen in the outflow driven by a protostar.
  Magnetic field lines are strongly twisted near the protoplanet and the outflows have well-collimated structures in this model.
 These outflows can be  landmarks for searching exo-protoplanets in their formation stages.
 Our results indicate that the accretion rate onto the protoplanet tend to have a larger value than that expected from previous hydrodynamical calculations, since a fraction of the angular momentum of circum-planetary disk is removed by outflows, enhanced non-axisymmetric patterns caused by magnetic field, and magnetic braking.
Possible implications for observation are also briefly discussed.
 
\end{abstract}


\section{Introduction}
 Can the giant gas protoplanets drive outflows through their formation process ? 
 Outflows are ubiquitous in star-forming region and  believed to be a universal phenomenon in star formation process.
 Recently, an outflow driven by a young brown dwarf was found in $\rho$ Ophiuchi star-forming region (Whelan \etal 2005).
 Recent numerical simulations show that protostellar outflows can be driven in the collapsing clouds when the clouds are initially magnetized and rotating \citep{tomisaka02,matsumoto04,machida04,machida05b,banerjee06}.
 In the planet formation process, the angular momentum is acquired from shearing motion between the protoplanet and protoplanetary disk.
 Thus, outflows may be driven by the protoplanet, when the protoplanetary disk is magnetized 
\citep{Sano2000,Salmeron2005,inutsuka05,Matsumura2006}. 
Tidal interaction between the protoplanet and the protoplanetary disk tends to open up a gap within the protoplanetary disk where the density is so small that sufficient ionization is available even only by cosmic rays.
 The possibility to launch outflow from the close vicinity of a protoplanet is investigated by \citet{fendt03}.
 In his analytical study, he assumes the mass accretion rate and angular velocity of the protoplanet according to the recent results of hydrodynamical simulations, and indicates that the protoplanet can drive outflow with $\gtrsim 60$ km\,s$^{-1}$ of the velocity if circum-planetary disk is magnetized.
 However, the magnetic field also affects the efficiency of angular momentum transfer within and from the circum-planetary disk, and hence changes the mass accretion rate.
 Thus, three-dimensional MHD simulations are needed to clarify whether the proto-planet drive outflows or not.

 The giant gas planet formation has been well investigated by one-dimensional hydrodynamical simulations.
 Those studies suggest that the giant planets are formed by the gradual buildup of a rocky core through coagulation of solids, and this core undergoes significant gas accretion once the core mass reaches $\simeq 10$ Earth mass (Mizuno 1980; Bodenheimer \& Pollack 1986; Ikoma \etal 2000).
 \citet{tanigawa02a} have studied the gravitational interaction between a protoplanet and an isothermal gaseous disk using their two-dimensional local hydrodynamical simulation with shearing sheet model, and shown the gas flow pattern around the protoplanet. 
  \citet{miyoshi99} and \citet{tanigawa02b} performed three-dimensional local hydrodynamical simulation and found the detailed flow pattern around the protoplanet.
 Pioneering work on the effect of the magnetic field in protoplanet formation was done by \citet{nelson03}.
 They have shown that accretion rate onto forming gas giant planet and flow pattern are modified by magnetic connection between the circumstellar disk and circum-planetary disk.

 In this paper, we calculate the formation of the gas giant protoplanet in the magnetized protoplanetary disk using three-dimensional nested grid method.
 We adopt the shearing sheet (i.e., local) model, and found the gas giant protoplanets can drive outflows.

\section{Model and Numerical Method}
 We consider only a local region around the protoplanet using shearing sheet model \citep[e.g.,][]{goldreich65}.
 We assume that the temperature is constant and the self-gravity of the disk is negligible in this local region.
 The orbit of the protoplanet is assumed to be circular on the equatorial plane of the disk.
 In our models, we set up local Cartesian coordinates with origin at the protoplanet's position and the $x$-, $y$-, and $z$-axis are, respectively, radial, azimuthal, and vertical direction of the disk.
 We solve ideal MHD equation without self-gravity:
\begin{eqnarray}
& \dfrac{\partial \rho}{\partial t} + \nabla \cdot (\rho \, \vect{v}) = 0,\\
& \dfrac{\partial \vect{v}}{\partial t} + (\vect{v} \cdot \nabla) \vect{v} =
 - \dfrac{1}{\rho} \nabla P - \nabla \psi_{\rm eff} - 2 \Omega_{\rm p} (\hat{\vect{z}} 
\times \vect{v}),  \\ 
& \dfrac{\partial \vect{B}}{\partial t} = \nabla \times (\vect{v} \times \vect{B}),
\end{eqnarray}
where $\rho$, $\vect{v}$, $P$, $\psi_{\rm eff}$, $\Omega_{\rm p}$, $\hat{\vect{z}}$, and $B$ are the gas density, velocity, gas pressure, effective potential, Keplerian angular velocity of the protoplanet, a unit vector directed to the $z$-axis, and the magnetic flux density, respectively.
   In the above equations, the curvature terms are neglected.
We adopt an isothermal equation of state, $P = c_{\rm s}^2 \rho$, 
 where $c_{\rm s}$ is sound speed. 
Because we neglect the self-gravity of the disk in equation~(2), 
 density is scalable in basic time evolution equations~(1),(2), and (3). 
Therefore we present our result for density distribution in units 
 of the initial density at the mid-plane of 
 the protoplanetary disk. 
   The Keplerian angular velocity of the protoplanet is given by
$\Omega_{\rm p} = ( G\, M_{\rm c}/ r_{\rm p}^3 )^{1/2}$,
where $G$, $M_{\rm c}$, and $r_{\rm p}$ are the gravitational constant, mass of the central star, and the distance between the protoplanet and the central star, respectively.
 Our calculations are normalized by unit time, $\Omega_{\rm p}^{-1}$, unit velocity, $c_{\rm s}$, and unit length,  $h \equiv c_{\rm s}/\Omega_{\rm P}$.
 The unit length corresponds to the scale height of the disk.
 The effective potential $\psi_{\rm eff}$ in our normalization is given by
$\psi_{\rm eff} = - (3 x^2 - z^2)/2 \,
 - \, 3\, \rh/(r^2+\epsilon^2)^{1/2}$.
 The first term is composed of the gravitational potential of the central star and the centrifugal potential, and higher orders in $x$, $y$ and $z$ are neglected.
 The second term is the gravitational potential of the protoplanet, where $\rh$, $r$, and $\varepsilon$ are Hill radius, the distance from the center of the protoplanet and softening parameter. The Hill radius are defined by
$
 \rh = ( 
M_{\rm p}/3M_{\rm c})^{1/3} r_{\rm p},
$ 
 where $M_{\rm p}$ is the mass of the protoplanet.
 In the unmagnetized disk, gas flow is characterized by only one parameter, $r_{\rm H}$. 
In this paper, we adopt $\rh = h$.  
This corresponds to the actual mass of protoplanet related to 
 the orbital radius and the mass of the central star as 
 $M_p = 3 M_{\rm c} (h/r_{\rm p})^{3} $ $\msun$.
In the standard solar nebular \citep{hayashi85}, the scale height of the disk at $r_{\rm p}= 5.2$ AU is  $h = 0.262$ AU.
 Thus, the mass of the protoplanet is $M_p = 0.4 M_J$ for $M_c = 1 \msun$,
  where $M_J$ is the mass of Jupiter.
We will discuss our results assuming 
 $r_{\rm p} = 5.2$ AU and $M_{\rm c} = 1\msun$ 
 in the following.
We will show the results with other sets of parameters 
 in a subsequent paper.
 The gas flow has a constant shear in the $x$-direction as
$
\vect{v_0} = ( 0,\,  -3/2\, x, \,0 ).
$
 The density is given by 
$
\rho = \sigma_0\, {\rm exp } ( - z^2/2h^2)/ (2\pi h^2)^{1/2},
$
 where $\sigma_0$ ($\equiv \int_{-\infty}^{\infty} \rho \, dz $) is the surface density of the unperturbed disk.
 The density is normalized by $\rho_0 = \sigma_0/h$  \citep[for detail see][]{miyoshi99}.
  We calculate three models: one unmagnetized disk model (Model B0), and two magnetized disk models (Models BZ and BY). 
 In model BZ, the magnetic field is perpendicular to the disk with uniform field strength as 
$\vect{B} = (0, \,0, \,0.1\,B_{\rm c})$,  where we adopt central magnetic field  as $B_{\rm c} = (8 \pi c_{\rm s}^2 \rho_0)^{1/2}$. 
 On the other hand, the magnetic field in Model BY is parallel to the azimuthal direction of the disk and decrease with increasing $z$ as
$
\vect{B} =(0,\, B_{\rm c}\, {\rm exp} [-z^2/2h^2], \,0 )$.
 In this paper, we show only strongly magnetized disk models in which the magnetic field energies are comparable to the thermal one, in order to see the magnetic effect.

  We adopt the nested grid method  \citep[for detail, see][]{machida05a} to obtain high spatial resolution near the origin.
  Each level of rectangular grid has the same number of cells ($ = 64 \times 128 \times 32 $),  but cell width $\Delta s(l)$ depends on the grid level $l$.
 The cell width is reduced 1/2 with increasing  grid level ($l \rightarrow l+1$).
 We use 6 grid levels ($l=$1,2$\cdot \cdot \cdot 6$).
 The box size of the  coarsest grid $l=1$ is chosen to $(x,y,z) = (16h, 32h, 8h)$, and that of  the finest grid is $(x,y,z) = (0.5h, h, 0.25h)$.
 The cell width of the coarsest grid is $\Delta s(1) = 0.25 h$, while that of the finest grid has  $\Delta s(6)= 7.8\times10^{-3}h$. 
 We assume the fixed boundary condition in the $x$- and $z$-direction and periodic boundary condition in the $y$-direction.
 We adopt softening length as $\varepsilon = 0.03 h$ to mimic the object with a finite size.

\section{Results}
Figure~\ref{fig:1} shows the density distributions and velocity vectors at $t=6.11$ for model B0.
We overplot 5 or 3 levels of  grids.
The protoplanet revolves counterclockwise direction around the central star approximately once by this time.
In our calculations, the flow patterns in all models seem to become steady on the equatorial plane in $t\simeq 5$.
Although we calculate the evolution until $t \simeq 10$, the flow patterns hardly change during $ 5\lesssim t \lesssim 10$.
 In Figure~\ref{fig:1} {\it a}, the gas flows according to the Keplerian shear motion outside the Hill radius ($r > 1$): gas enters from upper $y$ boundary for $x>0$ (from lower $y$ boundary for $x<0$) and goes downward (upward for $x<0$). 
 In the region of $\vert x \vert \gtrsim r_{\rm H}$, the flow is directed to the protoplanet, and then the gas is confronted by a shock wave as shown in Figure~\ref{fig:1} {\it a}.
 Figure~\ref{fig:1} {\it b} is enlargement of Figure~\ref{fig:1} {\it a}.
 In this panel, the gas flow is bent due to shock wave.
 Then, gas turns round by the Coriolis force and goes upward in the region of $x < 0$.
  When the post-shocked gas flows into $r \lesssim 0.5$, the gas is accreted to the circumplanetary disk.
  The disk rotates circularly around protoplanet in the counterclockwise direction near the protoplanet ($r \lesssim 0.5$).
 In Figure~\ref{fig:1} {\it c}, the shock wave is observed on the vertical plane.
 Once the gas is vertically rolling up, it falls to the center inside the shock region.

Figure~\ref{fig:2} shows the density and velocity distributions for model BZ.
Although the large scale structure in the $z=0$ plane for model BZ (Fig.~\ref{fig:2} {\rm a}) are similar to that for model B0 (Fig.~\ref{fig:1} {\it a}), ellipsoidal density gaps appear at ($x$, $y$) $\simeq$ ($\pm 1.5$, $\mp 3$) that correspond to the low beta (magnetically dominated) regions.
 In Figure~2{\it a} and 2{\it b}, the regions enclosed by the red contour lines indicate magnetically dominated region ($\beta_p < 1$), where $\beta_p$ denotes $\beta_p \equiv 8\pi c_s^2 \rho / B^2$.
 In this region,  the magnetic pressure exceeds the thermal pressure, while the thermal pressure  dominates  outside this region. 
  The low density regions appear also inside the Hill radius ($r<1$) that also correspond to the low beta regions (Fig.~\ref{fig:2} {\it b}). 
In the neighborhood of protoplanet ($r \lesssim 0.5$), the gas enters from lower left and upper right and flow out to lower right and upper left, as seen in Figure~\ref{fig:2} {\it b}.
Thus, flow patterns in model BZ are  considerably different from those in model B0 near the protoplanet.
The structure of vertical direction in model BZ is also considerably different from that in model B0. 
Outflow has appeared in model BZ since $t \simeq 0.66$, while outflow has not appeared in model B0.
 White thick line in Figure~\ref{fig:2} {\it c} and {\it d} mean the boundary between the outflow and inflow (contour curve of $v_z$=0).
 We can see from Figure~\ref{fig:2} {\it c} and {\it d} that outflow speed is much faster than  the infall speed. 
 Therefore, gas density decrease steeply along the $z$-axis as shown in Figure~\ref{fig:2} {\it c} and {\it d}, because strong outflow vertically blows off the gas.
 The outflow took the maximum speed of $v_{\rm out} = 14.1$  at $t=3.37$.
Then, outflow became roughly steady state with velocity of $v_{\rm out} \simeq 10$ and gradually increases again in stages near the end of the calculation.
 Magnetic field lines are strongly twisted near the protoplanet, and outflow has a well-collimated (or cone-like) structure that is similar to the outflow driven by a protostar \citep{tomisaka02}.

An outflow also appears in model BY.
Cut planes for model BY are shown in Figure~\ref{fig:3}. 
Comparing Figure~\ref{fig:3} {\it a} with Figure~\ref{fig:1} {\it a}, the gas flow in model BY is similar to that in model B0 on the large-scale.
 However, the flows in small-scale are different as indicated by Figure~\ref{fig:3}{\it b} and \ref{fig:1}{\it b}. 
Spiral waves are appeared in model BY (Fig.~\ref{fig:3} {\it b}), while the central region has a round shape in model B0 (Fig.~\ref{fig:1} {\it b}). 
These patterns seem to be caused by magnetic force.
The magnetic and thermal pressure dominated regions are clearly separated from each other in this model. 
The gaps in Figure~\ref{fig:3} {\it a} and {\it b} correspond to the magnetic pressure dominated region as well as those of model BZ.
The gas flows out of this region because of the strong magnetic pressure gradient.
Outflow has appeared since $t \simeq 3.6$ in model BY.
Inside the white thick lines that correspond to the outflow region, we can see the density contours expand outside, which indicates the mass ejected from the vicinity of the protoplanet owing to the outflow.
The maximum speed of the outflow changes with time.
The outflow becomes weak at $t = 4.65$ and has a $v_{\rm out} = 6.03$ of the maximum outflow speed at this epoch.
Then outflow becomes strong again, and has  $v_{\rm max} = 12.8$ at the end of the calculation ($t = 13.4$).
We found that outflows have appeared just around low beta regions and the outflowing region has a hollow tube structure as discussed in \citet{fendt03}.

\section{Discussion}
As shown in Figure 2 and 3 the gas is flowing away from the Hill region despite the gravity from the protoplanet. 
The escape velocity at $r= \rh = 0.04 r_{\rm p} $ 
 is 1.61$\km$ which is slower than the escape velocity 
 from the central star 18.5$\km$ 
 in our settings.
The sound speed at the orbit of Jupiter (5.2 AU) corresponds to $c_{\rm s} = 0.66\km$ in the standard solar nebula \citep{hayashi85}.
Thus the outflow speeds at the end of the calculations correspond to 
 9.31$\km$ for model BZ and 8.45 $\km$ for model BY, 
 if the protoplanet resides at 5.2AU in our models. 
The escape velocity from the protoplanet (1.61$\km$) at the Hill radius 
 is slower than the outflow speed derived from the final snapshot 
 of our calculation.
As shown in Figure~\ref{fig:2} and \ref{fig:3}, 
 the outflow region expands far beyond the Hill radius ($r\gg1$). 
In a region far from the protoplanet, 
 outflow has the velocity much faster than the escape velocity 
 from the protoplanet.
If the outflow speed from the protoplanet exceeds the escape velocity 
 of the central protostar, gas is continuously accelerated and 
 eventually ejected into the ambient interstellar medium.
The escape velocity of the central star ($18.5\km$) at the orbit 
 of Jupiter is about two times faster than outflow speed derived 
 from the final snapshot of our calculations.
However, the outflow speed for model BZ continues to increase 
 until the end of our present calculation. 
Outflows may increase their speed with increasing mass of the protoplanet, 
 while the mass of the protoplanet is assumed to be fixed 
 in our calculations for simplicity. 
If the outflow driven by the protoplanet has the speed larger than 
 the escape velocity of the central star, we can observe it 
 analogously with the protostellar outflow.
Thus we will possibly infer the existence of the protoplanet 
 from the observed outflow driven by the protoplanet 
 with future high resolution observational facility such as ALMA.

\begin{figure}
\includegraphics[width=110mm]{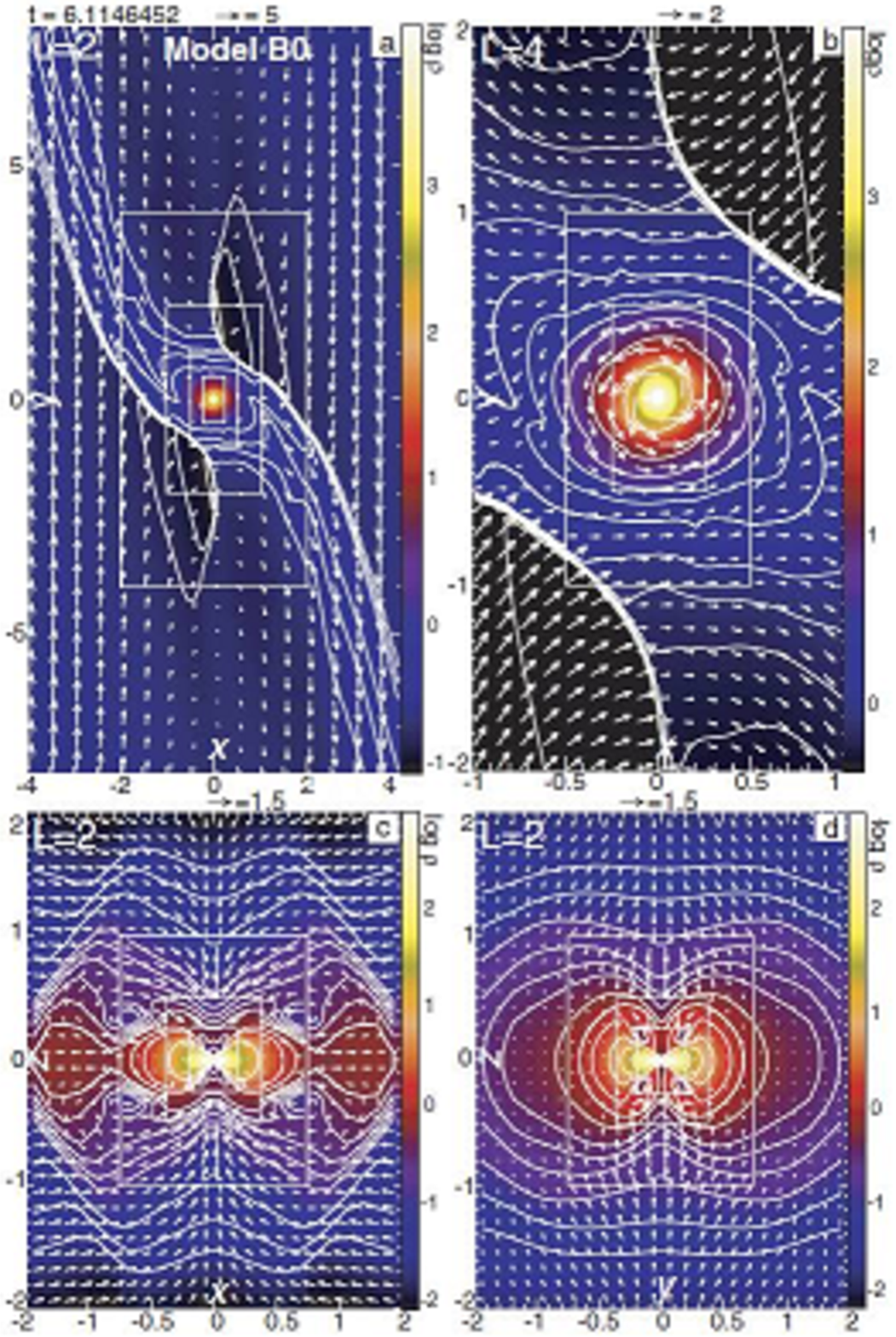}
\caption{
The density (color-scale and contour) and velocity distributions (arrows) on the cross-section in $z=0$ (equatorial plane; {\it a} and {\it b}), $y=0$ (radial direction; {\it c}), and $x=0$ plane (azimuthal direction; {\it d}) for model B0 are plotted at $t \simeq 6.11$.
Three levels of grid ($l=4,5,6$) are overplotted in panel {\it b}, while five levels of grid ($l=2,\cdot \cdot \cdot,6$) are oveplotted in panels {\it a}, {\it c} and {\it d}.
The scale of the velocity vectors, and  grid level ($l$) are shown in  each panel. 
The unit of the velocity (i.e., sound speed) corresponds to 0.66 $\km$, when we assume the orbit radius, gas temperature and mean molecular weight as $5.2$AU, 122.7 K, and 2.34, respectively.
}
\label{fig:1}
\end{figure}

\begin{figure}
\includegraphics[width=110mm]{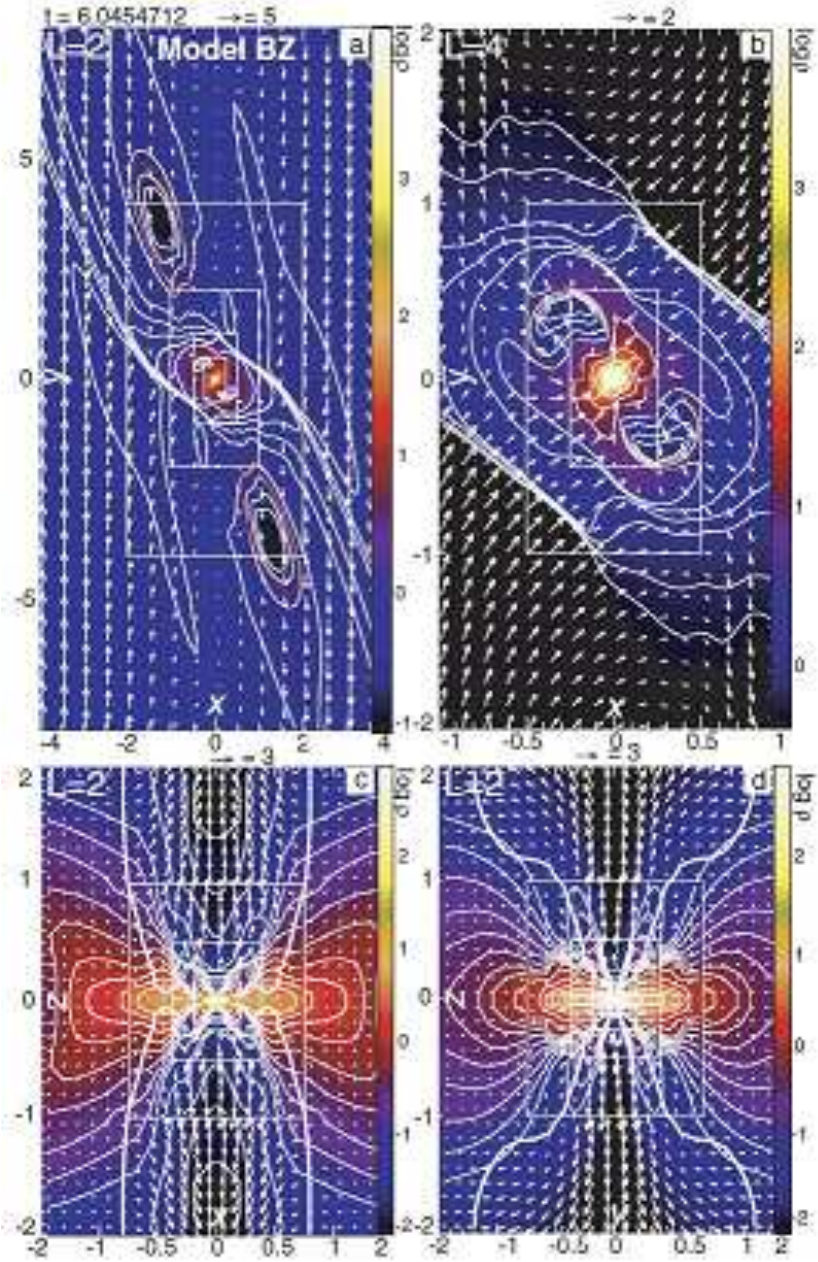}
\caption{
Same as Figure~\ref{fig:1} but for model BZ at $t \simeq 6.05$.
Thick red line in panels {\it a} and {\it b} means the magnetically dominated region ($\beta_{\rm p} < 1$, where $\beta_{\rm p}$ is plasma beta).
Thick white lines in panels {\it c} and {\it d} are the boundary between the inflow and outflow region (contour line of $v_z=0$).
Inside white line, gas is outflowing.
}
\label{fig:2}
\end{figure}

\begin{figure}
\includegraphics[width=110mm]{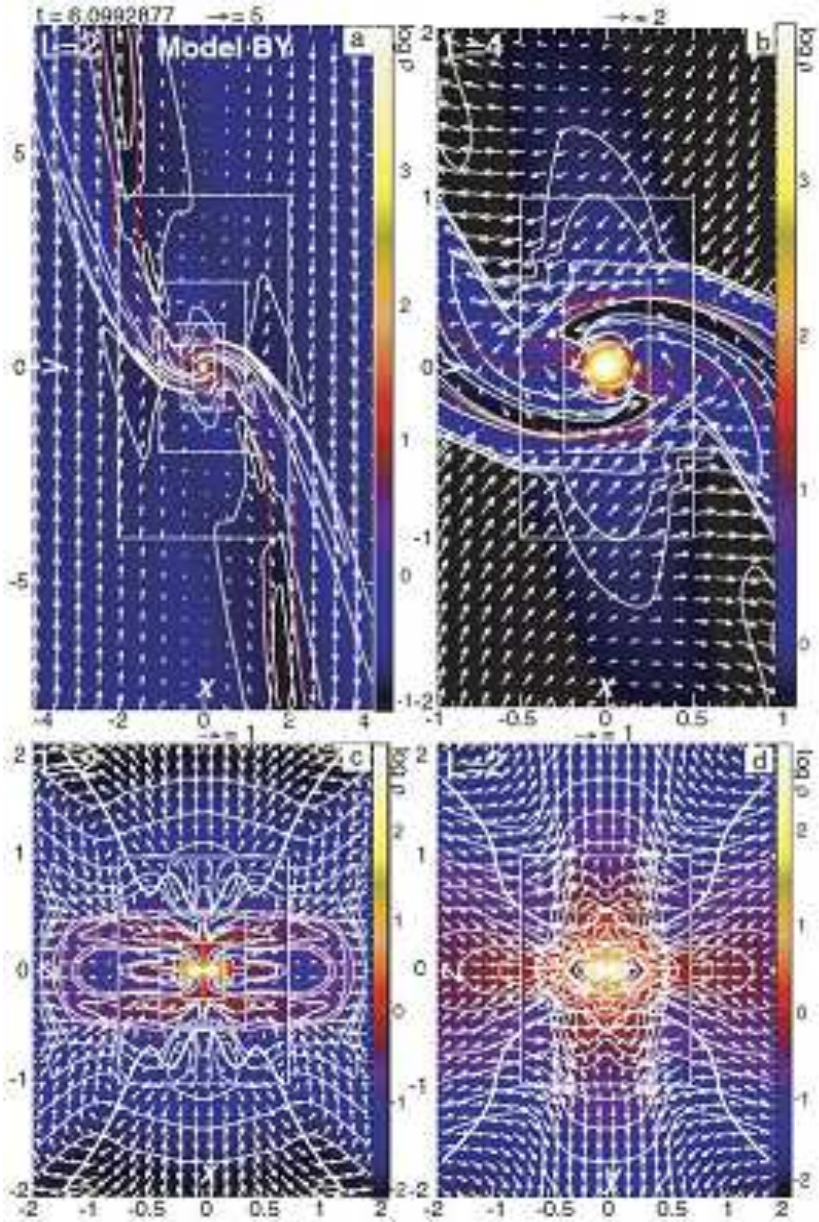}
\caption{
Same as Figure~\ref{fig:2} but for model BY at $t \simeq 6.1$.
}
\label{fig:3}
\end{figure}

Outflow may influence the evolution of the protoplanet.
Angular momentum is significantly transported by the outflow 
 as in the star-formation process \citep{tomisaka02}. 
Angular momentum is also transported by non-axisymmetric pattern 
 near the protoplanet.
In our calculations, the spiral patterns are formed near the 
 protoplanet ($r \lesssim 0.5$) in magnetized disk models as seen in
 Figure~\ref{fig:2} {\it b} and Figure~\ref{fig:3} {\it b},
 while no non-axisymmetric patterns appears within $r \lesssim 0.5$ 
 in unmagnetized disk model (Fig.~\ref{fig:1} {\it b}).
Furthermore, the effect of magnetic braking is important for 
 transportation of the angular momentum.
To illustrate this, 
 we measured the mass and angular momentum of the gas inside 
 $r<r_{\rm H}$ at the end of the calculation, 
 and found that 
 the acquired angular momenta per accreted mass
 are 0.083 (model B0), 0.058 (model BY), and 0.062 (model BZ) 
 in units of $(c_{\rm S}^2/\Omega_{\rm p})$.   
Note that we followed only later phase of the giant planet formation; 
 the initial mass of the protoplanet is already 0.4$\mj$.  
Nonetheless, our result shows that the angular momentum of the gas accreting onto the protoplanet is more effectively transferred outside in magnetized disk models. 
The evolution of the protoplanet's rotation should be studied  in long-term calculations.

\acknowledgments
We thank Drs.~H. Tanaka, and T. Tanigawa for valuable discussion,
and T. Hanawa for contribution to the numerical code. Numerical
calculations were carried out with a Fujitsu VPP5000 at ADAC of
NAO Japan. This work is supported by the Grant-in-Aid for the 21st 
Century COE "Center for Diversity and Universality in Physics" from 
the Ministry of Education, Culture, Sports, Science and Technology 
(MEXT) of Japan, and partially supported by the Grants-in-Aid from 
MEXT (15740118, 16077202, 16740115).

\newpage

\end{document}